\documentclass[pra,twocolumn,showpacs,preprintnumbers,amsmath,amssymb]{revtex4}
\usepackage{hyperref}
\usepackage{times}
\usepackage{soul,xspace}
\usepackage{graphicx}
\usepackage{hypernat}
\usepackage{ellipsis}
\usepackage[figure,table]{hypcap}

\hypersetup{colorlinks=true}

\mathchardef\ordinarycolon\mathcode`\:
\mathcode`\:=\string"8000
\begingroup \catcode`\:=\active
  \gdef:{\mathrel{\mathop\ordinarycolon}}
\endgroup

\newcommand{\di}{\ensuremath{\mathrm{d}}}
\newcommand{\im}{\ensuremath{\mathrm{i}}}
\newcommand{\eu}{\ensuremath{\mathrm{e}}}

\newcommand{\GPE}{\caps{GPE}\xspace}
\newcommand{\BEC}{\caps{BEC}\xspace}

\begin{document}
\title{Self-trapping of Bose-Einstein condensates expanding into shallow
optical lattices}
\author{Matthias Rosenkranz}
\email{m.rosenkranz@physics.ox.ac.uk}
\author{Dieter Jaksch}
\homepage{http://www.physics.ox.ac.uk/qubit/}
\affiliation{Clarendon Laboratory, University of Oxford, Parks Road, Oxford
OX1 3PU, United Kingdom}
\affiliation{Keble College, Parks Road, Oxford OX1 3PG, United Kingdom}
\author{Fong Yin Lim}
\author{Weizhu Bao}
\homepage{http://www.math.nus.edu.sg/~bao/}
\affiliation{Department of Mathematics and Center for Computational Science
and Engineering, National University of Singapore, Singapore 117543}
\date{\today}

\begin{abstract}
  We observe a sudden breakdown of the transport of a strongly repulsive
  Bose-Einstein condensate through a shallow optical lattice of finite width.
  We are able to attribute this behavior to the development of a self-trapped
  state by using accurate numerical methods and an analytical description in
  terms of nonlinear Bloch waves. The dependence of the breakdown on the
  lattice depth and the interaction strength is investigated. We show that it
  is possible to prohibit the self-trapping by applying a constant offset
  potential to the lattice region. Furthermore, we observe the disappearance of
  the self-trapped state after a finite time as a result of the revived
  expansion of the condensate through the lattice. This revived expansion is
  due to the finite width of the lattice.
\end{abstract}
\maketitle

\section{Introduction}

The transport properties of Bose-Einstein condensates (\BEC{}s)
through optical lattices have sparked interest in recent years after
a series of experiments revealed dissipative dynamics and
instabilities~\cite{BhaMadMor97,BurCatFor01,MorMueCriCiaAri01,ScoMarBuj04}.
Early experiments with \BEC{}s in optical lattices showed characteristic
effects of such a periodic potential on atoms, namely, Bloch
oscillations~\cite{BenPeiRei96} and Josephson
junctions~\cite{AndKas98,CatBurFor01}. In later years the research focus
shifted towards the study of nonlinear effects arising due to interaction of
the atoms~\cite{ChrMorMue02}. Theoretical work suggested that two mechanisms,
energetic and dynamical instabilities, would lead to a dissipative
dynamics~\cite{WuNiu01,WuNiu03}. This dissipative behavior was also observed
experimentally, and it was possible to separate dynamical and energetic
instabilities~\cite{BurCatFor01,FalSarLye04,SarFalLye05}. In a deep lattice it
was found that increasing the nonlinearity leads to a self-trapped state
within the lattice~\cite{AnkAlbGat05}. Such a self-trapping effect was
predicted theoretically in the limit of the tight-binding
model~\cite{TroSme01}. A generalized theoretical framework for the experiment
in Ref.~\cite{AnkAlbGat05} was derived in~\cite{AleOstKiv06}, where it was
pointed out that the experimental results can be explained in terms of a
certain type of self-trapped state.

\begin{figure}
  \includegraphics[width=.8\linewidth]{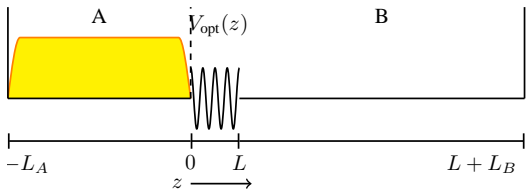}
  \caption{(Color online) Schematic setup of the system. The \BEC is initially
  located in the flat-bottom box reservoir A. The shutter to its right (dashed
  vertical line) can be removed instantaneously so that the atoms expand into
  the optical lattice $V_\text{opt}(z)$ of size $L$. A wide potential-free
  region B serves as a second reservoir for the atoms. In our numerical
  calculations we used the dimensionless lengths $L_A = 160\pi$, $L_B = 326\pi$
  and $L = 10\pi$ (for an explanation of the units see text).}\label{fig:setup}
\end{figure}

In this paper, we study the transport of a strongly interacting,
one-dimensional (1D) \BEC partially exposed to a shallow optical lattice of
finite width. A typical experiment in this field is conducted by trapping atoms
in a parabolic trap and then switching on a moving optical lattice.
Alternatively, one can displace the parabolic trap in a stationary lattice,
which leads to acceleration of the atoms through the lattice. We model these
experiments by a simplified setup where two flat-bottom potentials are
connected via the lattice as shown in Fig.~\ref{fig:setup}. Instead of a moving
lattice, we make use of the inherent expansion of the \BEC. In contrast to
experiments such as~\cite{AnkAlbGat05}, in this present work we focus on the
very shallow lattice regime, where the lattice depth is strictly less than the
photon recoil energy. The theoretical study in~\cite{AleOstKiv06} suggests that
self-trapping is possible even at lower lattice depths but it does not take
into account a short extent of the optical lattice. We extend previous works in
this field by assuming an optical lattice of finite width connecting two
reservoirs. Counterintuitively, even in the shallow lattice regime, we observe
self-trapping for sufficiently high interaction strength. In contrast to
spatially localized gap solitons~\cite{ZobPoeMey99}, this type of localization
extends over a few lattice sites without decay. The localization can be
destroyed by applying a constant offset potential in the lattice region. The
numerical analysis also shows that the finite width of the lattice results in
the dissipation of the self-trapping after a finite time. We further compare
the numerically obtained localized states with an analytical description. For
our analytical results we incorporate nonlinear Bloch waves to approximate the
self-trapped state. For the numerical treatment we simulate the 1D
Gross-Pitaevskii equation (\GPE), which is known to be a valid mean-field
description at zero temperature~\cite{DalGioPit99}. To study the dynamics of
the system we use a second-order time-splitting spectral method (\caps{TSSP})
to solve the \GPE. This method is explicit, unconditionally stable and
spectrally accurate in space. It is known to yield higher accuracy for the time
evolution of the 1D \GPE with less computational resources compared to, for
example, Crank-Nicholson methods~\cite{BaoJinMar02,BaoJakMar03}.

One of the underlying motivations of this field is a growing interest in the
quantum transport of atoms for future microdevices. For such a device to become
possible it will be necessary to identify basic building blocks---the
equivalents of resistors, capacitors, transistors etc. in electronic devices---
and to understand their interconnection~\cite{BonSen04}. One approach is to use
optical lattices to mimic the crystalline structure found in electronic
devices~\cite{SeaKraAnd07}. In classical electronic circuits such structures
are found in wires and many fundamental components such as transistors or
diodes. Recently, metallic behavior of ultracold atoms in three-dimensional
optical lattices has been observed experimentally~\cite{McKWhiPas08}. This
finding further substantiates the analogy between electronic and atomic
circuits. In an atomic circuit, atomic diodes~\cite{RusMug04,HuWuDai07} and
transistors~\cite{MicDalJak04} would be connected by atom-transporting wires to
build more complex devices. Such a wire can be constructed with a 1D optical
lattice, which defines a band structure similar to the bands found in metals.
In contrast to their electronic counterparts, the properties of atomic wires
can be dynamically tuned in a precisely controllable way. By changing the
angle, frequency or power of the lasers creating the lattice, the band
structure of the medium can be adjusted to the required characteristics.

In Fig.~\ref{fig:setup} we sketch our full setup. The flat-bottom reservoirs A
and B serve as a source and sink of a \BEC, respectively. The \BEC is initially
confined to a potential-free box A. Such a flat-bottom potential has been
implemented experimentally by Meyrath et al.~\cite{MeySchHan05}. The short
optical lattice connecting the reservoirs could be implemented by focusing two
laser beams very tightly. However, we stress that our setup is to be understood
rather as a generic theoretical model for a broader class of experimental
setups, for example with shallow harmonic oscillator potentials as reservoirs.
The difference in chemical potential on the left- and right-hand sides of this
lattice leads to an expansion of the atoms into the lattice and eventually into
reservoir B. In the atomic circuit picture, the setup can be seen as a battery,
and the expansion of the atoms leads to a discharge current.

The paper is organized as follows. In Sec.~\ref{sec:model} we will give a
detailed overview of the model used and introduce the quantities of interest.
Our numerical and analytical results of the transport properties of the \BEC
are explained in Sec.~\ref{sec:breakdown}. In the same section we also briefly
discuss the creation of solitons, which are generated in our scheme. We
conclude the paper with a summary in Sec.~\ref{sec:conclusion}. An appendix
provides details about the calculation of the nonlinear band structure of an
interacting \BEC in an optical lattice.

\section{The model\label{sec:model}}
We consider a \BEC at zero temperature in the elongated trap $V(x, y, z) =
\tfrac{1}{2}m [\omega_\perp^2 (x^2+y^2) + V_\text{ax}(z)]$, where $m$ is the
mass of an atom in the \BEC. The transversal frequency $\omega_\perp$ is chosen
to be $\hbar\omega_\perp \gg gn_0$, where $g$ the interaction strength of the
atoms and $n_0$ the peak density of the \BEC.
In this paper we assume $g > 0$, which corresponds to repulsive atomic
interaction. The interaction strength can be expressed in terms of the s-wave
scattering length $a_s$ as $g = 4\pi\hbar^2 a_s/m$. The above choice of the
frequency $\omega_\perp$ results in the freezing of the atomic motion in the
radial directions. Hence, we treat the \BEC as an effectively one-dimensional
condensate with a trapping potential $V_\text{ax}(z)$ along the axial
direction~\cite{Ols98,PetShlWal00}. In our model this potential has the form
\begin{equation}\label{eq:potential}
  V_\text{ax}(z) = \begin{cases}
    0 & \quad\text{for } -L_\text{A} \leq z < 0,\\
    V_\text{opt}(z) & \quad\text{for } 0 \leq z \leq L,\\
    0 & \quad\text{for } L < z \leq L+L_\text{B},
  \end{cases}
\end{equation}
which is illustrated in Fig.~\ref{fig:setup}. The condensate is initialized in
the ground state of the box potential of size $L_\text{A}$, which is obtained
by solving numerically the time-independent \GPE using the normalized gradient
flow method~\cite{BaoDu04,BaoCheLim06}. After the initialization of the \BEC in
reservoir A, the shutter confining the \BEC (dashed line in
Fig.~\ref{fig:setup}) is removed. The \BEC then penetrates a short optical
lattice $V_\text{opt}(z) = V_0 + V_1 \cos(2kz)$ of size $L \ll L_\text{A}$. We
checked numerically that possible discontinuities in $V_\text{ax}$ at $z=0$ and
$z=L$ do not affect the overall results. The periodicity $k$ is given by the
geometry and wave number of the lasers producing the standing wave, and it
determines the number of lattice sites $Lk/\pi$. The lattice height $V_1$ and
the constant bias $V_0$ are assumed to be independently adjustable. The size of
the sink reservoir B is $L_\text{B} \gg L$.

In order to obtain the dimensionless 1D \GPE we introduce the following
dimensionless quantities. Times are rescaled according to $\tilde t = t
2E_\text{R}/\hbar$ and lengths according to $\tilde z = kz$, where $E_\text{R}
= \hbar^2 k^2/2m$ is the photon recoil energy. Inside the lattice region the
potential Eq.~\eqref{eq:potential} is given by $v_\text{ax}(z) =
v_\text{opt}(z) = V_\text{opt}(\tilde z)/E_\text{R} = v + s\cos(2\tilde z)$ and
zero otherwise. Hence, the dimensionless constant offset is $v =
V_0/2E_\text{R}$ and the lattice depth $s = V_1/2E_\text{R}$. Furthermore, the
wave function yields $\tilde\psi(\tilde z, \tilde t) = k^{-1/2}\psi(z, t)$. At
$T=0$ the \BEC can then be described by the dimensionless 1D \GPE
\begin{equation}\label{eq:gpe}
  \im\partial_t\psi = \left[-\frac{1}{2}\partial_z^2 + v_\text{ax}(z) +
  \beta|\psi|^2\right]\psi,
\end{equation}
where the tildes have been removed for clarity. The dimensionless interaction
strength is $\beta = N_\text{tot} a_s k \hbar\omega_\perp/E_\text{R}$, which is
expressed in terms of the number of atoms $N_\text{tot}$ and the s-wave
scattering length $a_s$. The wave function $\psi = \psi(z, t)$ is normalized
according to $\int|\psi(z,t)|^2\di z = 1$ for all times $t$.

As an indicator of the dynamics of the system we define the dimensionless
current
\begin{equation}\label{eq:current}
  j(z, t) = \frac{1}{2\im} \left[\psi^*(z, t)\partial_z\psi(z, t) -
  \psi(z, t)\partial_z\psi^*(z, t)\right].
\end{equation}
As will become apparent in the numerical analysis, it is advantageous to also
define a more qualitative quantity, namely, the stationary current within the
lattice. We compute the stationary current by taking the time derivative of the
particle number in reservoir B, $N_\text{B}$, at times where the particle
number $N$ within the lattice is nearly constant. Given a time $t_0$ with such
a nearly constant particle number we define the stationary current as
\begin{equation}\label{eq:j0}
  j_0 = \left.\frac{\di N_\text{B}}{\di t}\right|_{t_0}.
\end{equation}
In general, the stationary current will depend on all parameters of the
system such as the lattice depth $s$ or the interaction strength $\beta$.

\section{Breakdown of the atomic expansion\label{sec:breakdown}}
\begin{figure}
  \includegraphics[width=.8\linewidth]{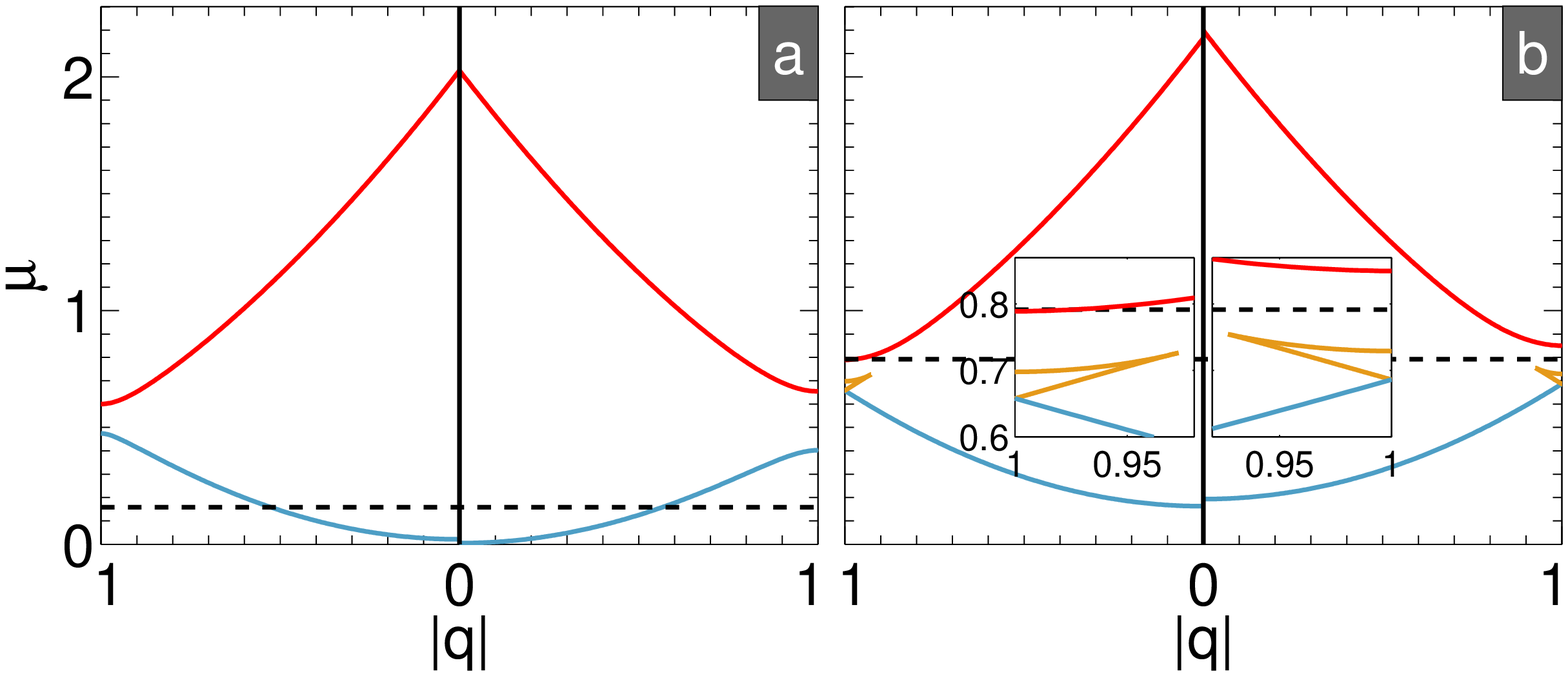}
  \caption{(Color online) Band structure for four sets of parameters $\mu$,
  $s$, and $n\beta$. The vertical line separates two sets with the same
  chemical potential (dashed line) but differing $s$ and $n\beta$. Note that
  each set is symmetric around $q=0$, hence we plot $|q|$. The offset $v$ is
  kept at zero. The other parameters are (a) $\mu = 0.16$ ($\beta = 79.58$)
  with $s = 0.13$, $n\beta = 0.05$ (left part) and $s = 0.25$, $n\beta = 0.04$
  (right part), (b) $\mu = 0.795$ ($\beta = 397.89$) with $s = 0.095$, $n\beta
  = 0.329$ (left) and $s = 0.127$, $n\beta = 0.393$ (right). The left and right
  insets in (b) show a zoom of the loops near the left and right band edges,
  respectively.}\label{fig:bands}
\end{figure}

Ignoring the finite width of the lattice, Eq.~\eqref{eq:gpe} reduces to to the
well-known Mathieu equation in the limit of vanishing interaction ($\beta =
0$). In this limit its energy eigenvalues develop the characteristic band
structure of periodic potentials. The inclusion of the nonlinear term
introduces a new energy scale $n\beta$ into the system ($n$ is the average
particle density within the lattice), which changes the band structure. A net
effect is an overall mean-field shift of the energies by $n\beta$. This effect
is shown in Fig.~\ref{fig:bands}(a) for typical parameters used in our
simulations. For $n\beta > s$ the band structure additionally develops a loop
at the band edge, which gradually decreases the width of the first band
gap~\cite{WuNiu00,DiaJenPet02,MacPetSmi03}. These loops can be observed in
Fig.~\ref{fig:bands}(b). The insets show a closeup of the band edges, where a
loop has developed. Note that we only plot $|q|$ in a reduced zone scheme,
which means that the loop closes symmetrically at $|q| > 1$. The relative
position of the chemical potential (dashed line in the figure) and the band gap
will be important for the explanation of the localized state in the next
subsection. For details about the calculation of the band structure we refer to
the appendix.

\subsection{Numerical results\label{ssec:numerics}}
\begin{figure}
  \includegraphics[width=.8\linewidth]{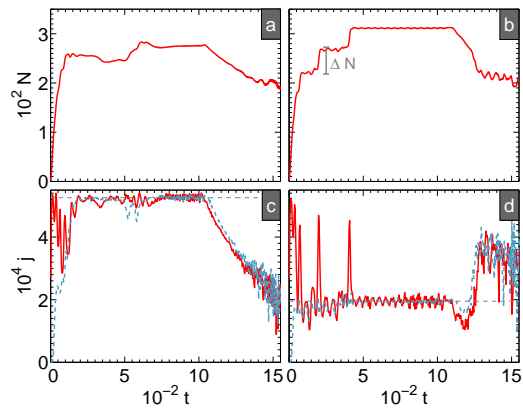}
  \caption{(Color online) Upper panel: time-dependent particle number within the
  lattice. Lower panel: time-dependent currents at the beginning of the lattice
  (red solid line) and the last lattice site (blue dashed). The gray dashed
  horizontal line indicates the stationary current obtained as described in the
  text. Lower panel: The parameters are $\beta = 397.89$, $v=0$ for all plots
  and (a), (c) $s = 0.095$ and (b), (d) $s = 0.127$. The vertical bar in (b)
  indicates the analytical result for the particle number difference $\Delta N$
  (see text).}\label{fig:current}
\end{figure}

In this section we will numerically investigate the transport of the BEC
initially trapped in region A through the lattice. We will connect the various
numerical findings and explain them in terms of self-trapped states with finite
life time.

We have calculated the stationary current Eq.~\eqref{eq:j0} numerically.
Figures~\ref{fig:current}(a) and (b) show two typical time-dependent plots of
the particle number within the lattice. We can recognize plateaux at different
times in Figs.~\ref{fig:current}(a) and \ref{fig:current}(b), which can be used
to compute the stationary current. For example, in Fig.~\ref{fig:current}(b)
such plateaux
exist at the three time intervals $[0.008, 0.019]$, $[0.021, 0.04]$, and $[0.04,
0.11]$. In the lower panel of Fig.~\ref{fig:current} two typical results for
the time-dependent current Eq.~\eqref{eq:current} are shown at different
positions within the lattice. The dashed horizontal line indicates the
stationary current corresponding to the same set of parameters. Note that its
value coincides well with the actual current within the lattice for an extended
amount of time. The current undergoes small oscillations around the value of
the stationary current. The stationary current indicates the gross expansion
speed of the \BEC. We will analyze its dependence on the parameters $s$ and
$\beta$ in the following.

\begin{figure}
  \includegraphics[width=.8\linewidth]{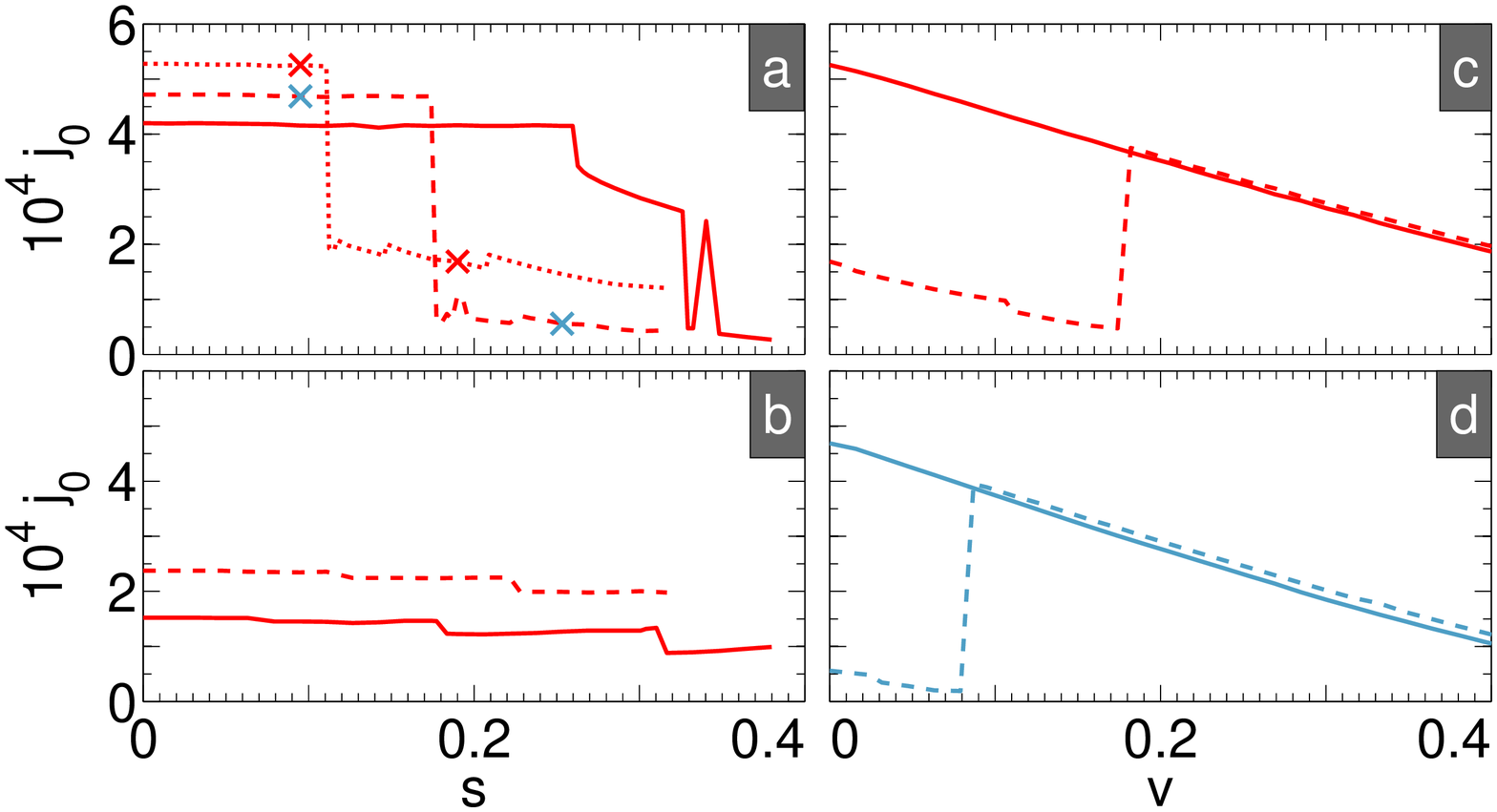}
  \caption{(Color online) Stationary current for (a), (b) varying $s$ at $v=0$
  and (b), (c) varying constant offset $v$ at fixed $s$. Parameters for (a) are
  $\beta=251.46$ (solid line), $\beta=318.31$ (dashed), and $\beta=397.89$
  (dotted), for (b) $\beta=31.83$ (solid) and $\beta=79.58$ (dashed). Plot (c)
  is at $\beta = 397.31$ for fixed $s = 0.095$ (solid line) and $s = 0.19$
  (dashed), (d) is at $\beta=318.31$ with $s=0.095$ (solid) and $s=0.253$
  (dashed). The values for the lattice amplitude $s$ used in (c), (d) are
  marked in (a) on the curves with the respective interaction
  strengths.}\label{fig:drops}
\end{figure}

Intuitively one expects the \BEC in the setup of Fig.~\ref{fig:setup} to expand
into the optical lattice where its transport properties are subjected to the
modified band structure discussed above. We conducted numerical calculations in
the regime of weak lattices ($s \ll 1$)
and strong interaction ($\beta \gg 1$). If we plot the stationary current for
different interaction strengths $\beta$ as a function of the optical lattice
amplitude $s$, we notice a sharp drop in the curves for large $\beta$. In
Fig.~\ref{fig:drops}(a) this drop is clearly visible, whereas for the lower
interactions in Fig.~\ref{fig:drops}(b) it is absent in the shallow lattice
regime. Note that the value of $j_0$ at $s=0$ increases with increasing
$\beta$. This behavior is expected because the higher repulsive interaction
leads to a higher potential difference between the reservoirs, which drives
more atoms through the lattice region. An increase of the lattice amplitude
does not influence the stationary current in Fig.~\ref{fig:drops}(a) at first,
instead it stays constant up to the drop. After the drop it decreases with
increasing lattice depth.

\begin{figure}
  \includegraphics[width=.8\linewidth]{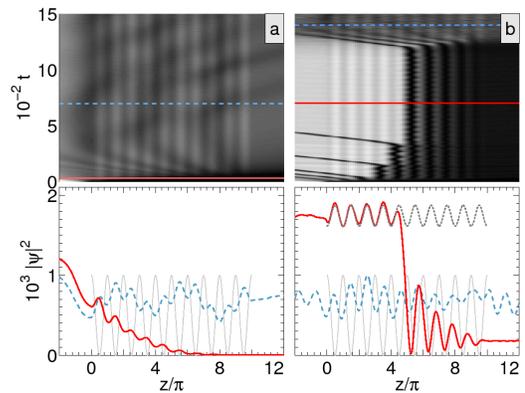}
  \caption{(Color online) Density plot of the \BEC near the optical lattice.
  The lattice extends over $0\leq z \leq 10\pi$. In the upper panel light
  shades indicate high density, dark shades low density. For a typical rubidium
  \BEC such as the one used in~\cite{AnkAlbGat05} the time scale of these plots
  is around $32\,\mathrm{ms}$, which is experimentally accessible. The lower
  panel shows the profile of the \BEC at the times indicated by the
  corresponding horizontal lines in the upper panel. The dotted line indicates
  the analytical result Eq.~\eqref{eq:trial2-expand} for the corresponding
  nonlinear Bloch wave and the thin solid gray line indicates the position of
  the optical lattice. For both plots $s = 0.127$, $v=0$ and (a) $\beta =
  79.58$, (b) $\beta = 397.89$.}\label{fig:density}
\end{figure}

We were able to relate the sudden drop in the stationary current to the
development of an extended plateau in the time-dependent particle density
within the lattice. In Fig.~\ref{fig:current}(b) we clearly recognize such a
plateau for times around $4\times 10^{-2}$ to $11\times 10^{-2}$, as well as
shorter plateaux at earlier times. For the parameters of this plot, $\beta =
397.89$ and $s = 0.127$, a drop in the stationary current has already occurred
(cf. Fig.~\ref{fig:drops}(a)). Also note that the drop can be observed in the
time-dependent current in Fig.~\ref{fig:current}(d), which is lower than the
current in Fig.~\ref{fig:current}(c). This means that after the drop, the \BEC
density within the lattice stays constant for an extended amount of time and
there is only a small residual current flowing through the lattice. The \BEC
has effectively stopped its expansion despite its high repulsion and despite
the lattice being very shallow. This fact can also be observed in a density
plot of $|\psi(z, t)|^2$. Figure~\ref{fig:density}(a) shows the density of a
\BEC with a low interaction strength which does not show a drop in the
stationary current. In contrast, the density for a higher $\beta$ which
features a drop in the stationary current is plotted in
Fig.~\ref{fig:density}(b). The value of $s$ in this plot was chosen larger than
the value for the drop (cf. Fig.~\ref{fig:drops}(a)). By comparing
Fig.~\ref{fig:density}(b) with the particle number in Fig.~\ref{fig:current}(b)
we also identify the steps in the plateau structure of the particle number as a
tunneling of the \BEC across lattice sites. Every time a lump of the \BEC
tunnels to the next lattice site, the average particle density increases by a
fixed amount $\Delta N$ until it reaches a final lattice site where the
expansion stops and a quasi-stationary state develops. For a fixed lattice
height the final lattice site is determined by the value of $n\beta$ and $s$.
This becomes clear by comparing the chemical potential in Fig.~\ref{fig:bands}
(dashed horizontal line) with the position of the first band gap for a given
set of parameters. For $\beta=79.58$, where the stationary current does not
exhibit a drop, the chemical potential lies deep in the first band (cf.
Fig.~\ref{fig:bands}(a)). The atoms can populate the first band when being
injected into the lattice region and move through the lattice. For very high
$\beta$ however, the chemical potential initially lies above the
first band gap (cf. Fig.~\ref{fig:bands}(b)). As we increase $s$, the local
chemical potential within the lattice $n\beta$ also increases slightly and
shifts the band structure in such a way that the overall chemical potential
$\mu$ falls into the first band gap. With its chemical potential lying inside
the gap the wave function cannot expand anymore because there are no states
available. This manifests itself in the development of the quasi-stationary
state in Fig.~\ref{fig:density}(b). Increasing $s$ even further broadens the
gap and the stationary current decreases steadily. We refer to this localized
state as self-trapped state because its development strongly depends on the
value of the nonlinearity $\beta$. We find the sudden onset of
quasi-stationarity within the lattice only for high $\beta$, whereas for lower
$\beta$ and large lattice amplitude the \BEC does not penetrate the lattice.
Alexander et al.~\cite{AleOstKiv06} showed that such a self-trapped ``gap
wave'' in a different setup remains stable. However, due to the shortness of
the lattice in our setup, there is still a small current flowing through the
lattice. This leakage eventually causes the localized state to disappear after
a finite time. In Fig.~\ref{fig:density}(b) this breakdown can be observed for
$t > 1300$, when the \BEC dissipates over the whole lattice. The phenomenon of
such a finite life-time of the localized state has also been observed by Wang
et al. in a different optical lattice setup~\cite{WanFuLiu06}.

We further studied the case of a fixed $s$ with a varying offset potential $v$
confined to the lattice region. The two curves in each of
Figs.~\ref{fig:drops}(c) and \ref{fig:drops}(d) correspond to fixed values of
$\beta$, respectively. The solid and dashed lines in each plot correspond to
two different values of $s$. The lattice depths are chosen to lie before and
after the drop in the stationary current. As expected, for a system whose
chemical potential is located below the first band gap at $v=0$, the stationary
current decreases with increasing offset potential (solid lines). In this case
the offset shifts the whole band structure by a constant value. However, if the
lattice depth is chosen such that the chemical potential lies in a band gap at
$v=0$, we observe a sudden jump in the stationary current (dashed lines). This
counterintuitive behavior, an increasing current for a higher potential
barrier, can be explained by again noting that the constant offset shifts the
band structure. Increasing the offset will eventually result in a band
structure where the chemical potential does not lie in a gap anymore. Thus the
jump occurs when the chemical potential rejoins a band.

\subsection{Nonlinear Bloch waves\label{ssec:Bloch}}
The \GPE~\eqref{eq:gpe} with a periodic potential leads to the well-known
Mathieu equation in the limit of vanishing interaction ($\beta = 0$) if we
assume a stationary state $\psi(z, t) = \exp(-\im \mu t) \psi_q(z)$.
Here, $\mu$ is the chemical potential. The chemical potential of a \BEC in the
strongly interacting limit can be determined by utilizing the Thomas-Fermi
approximation, which yields $\mu = \beta/L_A$. Our numerical calculations
of the chemical potential of the initial 1D \BEC are in good agreement with the
Thomas-Fermi approximation for the parameters used in this paper. The solutions
$\psi_q(z)$ are Bloch functions
$\psi_q(z) = \exp(\im q z) \sum_{\ell = -\infty}^{\ell=\infty}c_\ell \exp(\im
2\ell z)$. The parameter $q$ is the quasi-momentum of the condensate.
To model the resulting wave function of the interacting case we similarly
assume a Bloch function representation of the state. To simplify the analytical
model we further truncate the Bloch waves according to~\cite{MacPetSmi03}
\begin{equation}\label{eq:trial}
  \psi_q(z) = \sqrt{n} \eu^{\im q z} \left(c_0 + c_{-1} \eu^{-2\im z} + c_1
  \eu^{2\im z} \right).
\end{equation}
The density $n$ is defined as the averaged relative density of the \BEC within
the lattice
\begin{equation}\label{eq:n}
  n = \frac{N}{L}.
\end{equation}
The normalization of the full wave function $\psi(z)$ requires that $|c_0|^2 +
|c_{-1}|^2 + |c_1|^2 = 1$. By using this method it is also possible to recover
the band structure. For details of the calculation of the nonlinear band
structure we refer to the appendix.

To understand the localization of the state after the drop consider
Eq.~\eqref{eq:trial} with parameters $\sqrt n c_{-1} = \sqrt n c_1 =: d_1/2$
and $d_0 := \sqrt n c_0$. For the ground state with $q=0$ this results in the
state
\begin{equation}\label{eq:trial2}
  \psi(z) = d_0 + d_1 \cos(2z).
\end{equation}
The normalization condition has to be rewritten as
\begin{equation}\label{eq:trial2-normalisation}
  |d_0|^2 + \frac{1}{2} |d_1|^2 = n.
\end{equation}
The coefficients $d_0$ and $d_1$ can be determined by minimizing the energy
functional under this constraint. This minimization procedure yields the two
equations
\begin{subequations}\label{eq:d0d1}
  \begin{align}
  2\beta d_0^3 + d_0 (3\beta d_1^2 - 2\mu) + s d_1 &= 0,\\
  3 \beta d_1^3 + 4 d_1 (2 + 3\beta d_0^2 - \mu) + 4s d_0 &= 0,
\end{align}
\end{subequations}
which can be solved analytically. Their real solution describes a nonlinear
Bloch wave~\cite{BroCarDec01,AleOstKiv06}. For the case of $q=0$ this wave
function describes an oscillation with the period of the lattice around a
finite value. To see this we assume a real solution of Eqs.~\eqref{eq:d0d1} and
calculate the density $|\psi(z)|^2$ from Eq.~\eqref{eq:trial2} to yield
\begin{equation}\label{eq:trial2-expand}
  |\psi(z)|^2 = d_0^2 + \frac{d_1^2}{2} + 2d_0 d_1 \cos(2 z) +
  \frac{d_1^2}{2}\sin(4 z).
\end{equation}
The constant offset in this equation is the particle density $n$
given in Eq.~\eqref{eq:trial2-normalisation}. The oscillation with the double
period $4 z$ can be neglected since for our parameter range $d_1^2/2 \ll 2|d_0
d_1|$. In the simulations we observe that for $s$ above the threshold for the
drop in the stationary current, a quasi-stationary state develops.
Figure~\ref{fig:density}(b) shows the development of such a localized nonlinear
Bloch wave and its disappearance over time for high interaction strength. In
the lower panel of Fig.~\ref{fig:density}(b) we show this behavior at two time
slices. At early times the \BEC tunnels across lattice sites as it expands into
the lattice. At time $t \approx 700$ (solid line) we recognize the nonlinear
Bloch wave in the lower panel of Fig.~\ref{fig:density}(b). The dotted line
overlapping with the numerical curve is the analytical density
Eq.~\eqref{eq:trial2-expand} with the same parameters as the numerical result.
We used the reduced chemical potential of the \BEC still left in reservoir A
and the lattice. We note that our analytical description of a nonlinear Bloch
wave coincides well with the numerical result. At a later time $t \approx 1400$
(dashed line) the \BEC is spread uniformly with a periodic modulation. In
contrast, for values of $s$ and $\beta$ where the stationary current does not
show a drop, the \BEC spreads uniformly with time. The lattice region only
slightly modulates the otherwise uniformly distributed atom density. In the
lower panel of Fig.~\ref{fig:density}(a) we see that at $t \approx 30$ (solid
line) the \BEC has expanded into the optical lattice but the density is only
slightly modulated with the period of the lattice. Similarly, at the later time
$t \approx 700$ (dashed line) the now uniformly spread \BEC is only slightly
modulated by the periodic potential. A quasi-stationary nonlinear Bloch wave
has not developed.

We further calculated the size of the steps between the plateaux in
the particle number plot in Fig.~\ref{fig:current}(b). Integrating
the particle density Eq.~\eqref{eq:trial2-normalisation} over one
lattice site yields
\begin{equation}\label{eq:deltaN}
  \Delta N = \pi \left(|d_0|^2 + \frac{1}{2}|d_1|^2\right).
\end{equation}
When the \BEC advances one lattice site the particle number within the lattice
should grow by $\Delta N$. The vertical bar in Fig.~\ref{fig:current}(b)
indicates the analytically calculated difference Eq.~\eqref{eq:deltaN} for the
parameters used in the plot with a reduced chemical potential as discussed
above. This result agrees with the difference of the numerically obtained
particle number plateaux. Given the above agreements of analytical and
numerical results we conclude that indeed the formation of a nonlinear Bloch
wave causes the breakdown of the stationary current.

\subsection{Dark solitons}
The \GPE supports soliton solutions for nonzero interaction $\beta$. These
solutions are shape-preserving notches or peaks in the density which do not
disperse over time. In the case of repulsive interaction without an optical
lattice, solitons are typically of the dark type (notches in the
density)~\cite{BurBonDet99,CarBraBur01} but both dark and bright solitons can
exist in \BEC{}s in optical lattices~\cite{AlfKonSal02,StrGutPar02}. Our
numerical results of the condensate density in Fig.~\ref{fig:density}(b) show
the creation of moving dark solitons when the condensate jumps to a neighboring
lattice site. In Fig.~\ref{fig:density}(b) we can see dark notches moving to
the left, away from the lattice region. These excitations move slower than the
local speed of sound ($c = \sqrt{\beta|\psi|^2}$) and do not change their shape
considerably over the simulation time. We also observed other typical features
of solitons such as the repulsion of two solitons approaching each other or the
phase shift across a soliton. Furthermore, we notice that solitons emit sound
waves traveling at the speed of sound. This happens when the center of mass of
a soliton enters a region of different mean density, which causes a change
in speed. A detailed analysis of soliton trajectories and their deformations in
a nonuniform potential has been presented in~\cite{ParProLea03}. It should be
noted that the creation of moving solitons in our simulations shows
similarities to the creation of solitonlike structures through
self-interference of the \BEC in hard-wall potentials~\cite{RusKneSch01}. This
self-interference could be caused by a small fraction of the \BEC being
reflected when the larger fraction tunnels through a peak of the optical
lattice. Furthermore, it is worth noting that the exact dynamics and stability
of the solitons also depends on the ratio
$\mu/\omega_\perp~$\cite{MurShlErt02}. In our simulations we did not take into
account the radial confinement, assuming that it is very tight and the overall
\BEC dynamics can be described by a quasi-1D model. We did hence not undertake
a thorough analysis of the soliton dynamics in our system as their creation can
be considered a side product of the development of the quasi-stationary state
in the lattice, which was the main focus of this work.

\section{Conclusion\label{sec:conclusion}}
We have investigated the effects of a finite width lattice on the
transport properties of a strongly interacting \BEC. To this end we
numerically solved the corresponding 1D \GPE and extracted relevant
quantities such as the atomic current and density. We also compared
the numerical with analytical results in terms of nonlinear Bloch
waves.

We found that even for low lattice depths a quasi-stationary state develops
after an initial expansion of the \BEC into the lattice. This results in a
sharp drop of the current in the lattice when the lattice depth and interaction
reaches a critical value. However, due to the finite extent of the lattice the
atoms can tunnel out of this state, which eventually leads to the breakdown of
the stationarity. We could explain the development of a stationary state with
partial nonlinear Bloch waves, which builds up over only a few lattice sites
and blocks further atom flow. When we introduced a constant offset potential
into the system, we found that increasing the offset can trigger the previously
suppressed flow of the atoms again. Finally, we reported on the creation of
moving dark solitons during the development of the nonlinear Bloch wave. Every
time the \BEC moves to a neighboring lattice site a soliton is emitted. Hence,
the number of solitons present in the \BEC indicates the number of occupied
lattice sites.

In the context of atomic circuits the present work illuminates the role of
wires in an atomic circuit. Our results suggest that by slightly increasing the
lattice depth the superfluid current through the wire can be stopped easily.
However, this effect is not based on a sudden Mott insulator transition since
we always assumed the applicability of the \GPE. In fact, it is based on a
self-trapped macroscopic configuration, where the ``charges'' penetrate the
wire to a certain depth but do not flow further.

\begin{acknowledgments}
  The authors acknowledge support from the Institute for Mathematical Sciences
  of the National University of Singapore, where parts of this work have been
  completed. This research was supported by the European Commission under the
  Marie Curie Programme through \caps{QIPEST} and Singapore Ministry of
  Education grant No. R-158-000-002-112.
\end{acknowledgments}

\appendix*

\section{Nonlinear band structure\label{sec:bands}}
As discussed in Sec.~\ref{ssec:Bloch} we assume a wave function of the form
Eq.~\eqref{eq:trial}. This trial function can be used to determine the energy
spectrum of the system. In the noninteracting case this leads to the
well-known linear band structure. The inclusion of the nonlinear term
$\beta|\psi|^2$ with $\beta > 0$ in the \GPE modifies the band structure of the
Mathieu eigenvalues. The overall mean-field shift of the band structure and the
development of loops have been discussed in Sec.~\ref{sec:breakdown}. Their
occurrence can be observed by assuming a trial function Eq.~\eqref{eq:trial}.
We then rewrite the normalization condition of the coefficients $c_\ell$ in
terms of two angles $\phi$ and $\theta$ according to
\begin{subequations}\label{eq:subst}
\begin{align}
  c_0 &= \cos\theta,\\
  c_{-1} &= \sin\theta\sin\phi,\\
  c_1 &= \sin\theta\cos\phi.
\end{align}
\end{subequations}

Stationary states can be found by plugging the ansatz Eq.~\eqref{eq:trial}
together with Eqs.~\eqref{eq:subst} into the energy functional
\begin{equation}\label{eq:energyint}
  \begin{split}
  \frac{\epsilon_q(\phi, \theta)}{n} &= \frac{1}{n\pi} \int_0^\pi
  \biggl(\frac{1}{2}|\nabla\psi_q(z)|^2 + s\cos(2z) |\psi_q(z)|^2\\
  &\phantom{=} + \frac{\beta}{2}|\psi_q(z)|^4 \biggr) \di z.
  \end{split}
\end{equation}
This integral is solved analytically and yields
\begin{equation}\label{eq:energy}
  \begin{split}
    \frac{\epsilon_q(\phi, \theta)}{n} &= \frac{q^2}{2} +
    2\sin^2\theta\,[1 + q \cos(2\phi)]\\
    &\hphantom{=} + \frac{s}{2}\sin(2\theta)\, (cos\phi + \sin\phi)\\
  &\hphantom{=} + \frac{n\beta}{64} \Bigl\{43 - \cos(4\phi)[3 + \cos(4\phi)]\\
  &\hphantom{=}- \cos(4\theta) [7 + 8\sin(2\phi)]\\
  &\hphantom{=}+ 8\sin(2\phi)[1 - \cos(2\theta) \sin(2\phi)]\Bigr\}.
  \end{split}
\end{equation}
The first line represents the kinetic energy, the second line the lattice
potential and the last three lines the interaction energy. By fixing different
values of the quasi-momentum $q$ and minimizing Eq.~\eqref{eq:energy} with
respect to $\phi$ and $\theta$ we recover the band structure for given
parameters $n\beta$ and $s$ (cf. Fig.~\ref{fig:bands}).
\vspace{.8cm}

\bibliography{lattice-trapping}

\end{document}